\documentclass[sigconf, nonacm]{acmart}

\AtBeginDocument{%
  \providecommand\BibTeX{{%
    \normalfont B\kern-0.5em{\scshape i\kern-0.25em b}\kern-0.8em\TeX}}}

\setcopyright{acmlicensed}
\copyrightyear{2025}
\acmYear{2025}
\acmDOI{XXXXXXX.XXXXXXX}

\acmConference[ SIGGRAPH 2025 ]{SIGGRAPH 2025}{August 10--14,
  2025}{Vancouver, CAN}
\acmISBN{978-1-4503-XXXX-X/18/06}

\acmSubmissionID{042}

\begin{document}

\title{Learnable Fractal Flames}
\subtitle{A Differentiable Approach to Image-Guided Fractal Art Synthesis}
\author{Jordan J. Bannister}
\email{jordan.bannister@mila.quebec}
\affiliation{%
  \institution{Mila - Quebec AI Institute}
  \streetaddress{6666 St-Urbain}
  \city{Montréal}
  \state{QC}
  \country{Canada}
  \postcode{H2S 3H1}
}

\author{Derek Nowrouzezahrai}
\email{derek@cim.mcgill.ca}
\affiliation{%
  \institution{McGill University}
  \streetaddress{845 Rue Sherbrooke Ouest}
  \city{Montréal}
  \state{QC}
  \country{Canada}
  \postcode{H3A 0G4}
  }

\renewcommand{\shortauthors}{Bannister and Nowrouzezahrai}

\begin{abstract}
This work presents a differentiable rendering approach that allows latent fractal flame parameters to be learned from image supervision using gradient descent optimization. The approach extends the state-of-the-art in differentiable iterated function system fractal rendering through support for color images, non-linear generator functions, and multi-fractal compositions. With this approach, artists can use reference images to quickly and intuitively control the creation of fractals. We describe the approach and conduct a series of experiments exploring its use, culminating in the creation of complex and colorful fractal artwork based on famous paintings.  
\end{abstract}



\received{XXX}
\received[revised]{XXX}
\received[accepted]{XXX}

\begin{teaserfigure}
    \includegraphics[width=\textwidth]{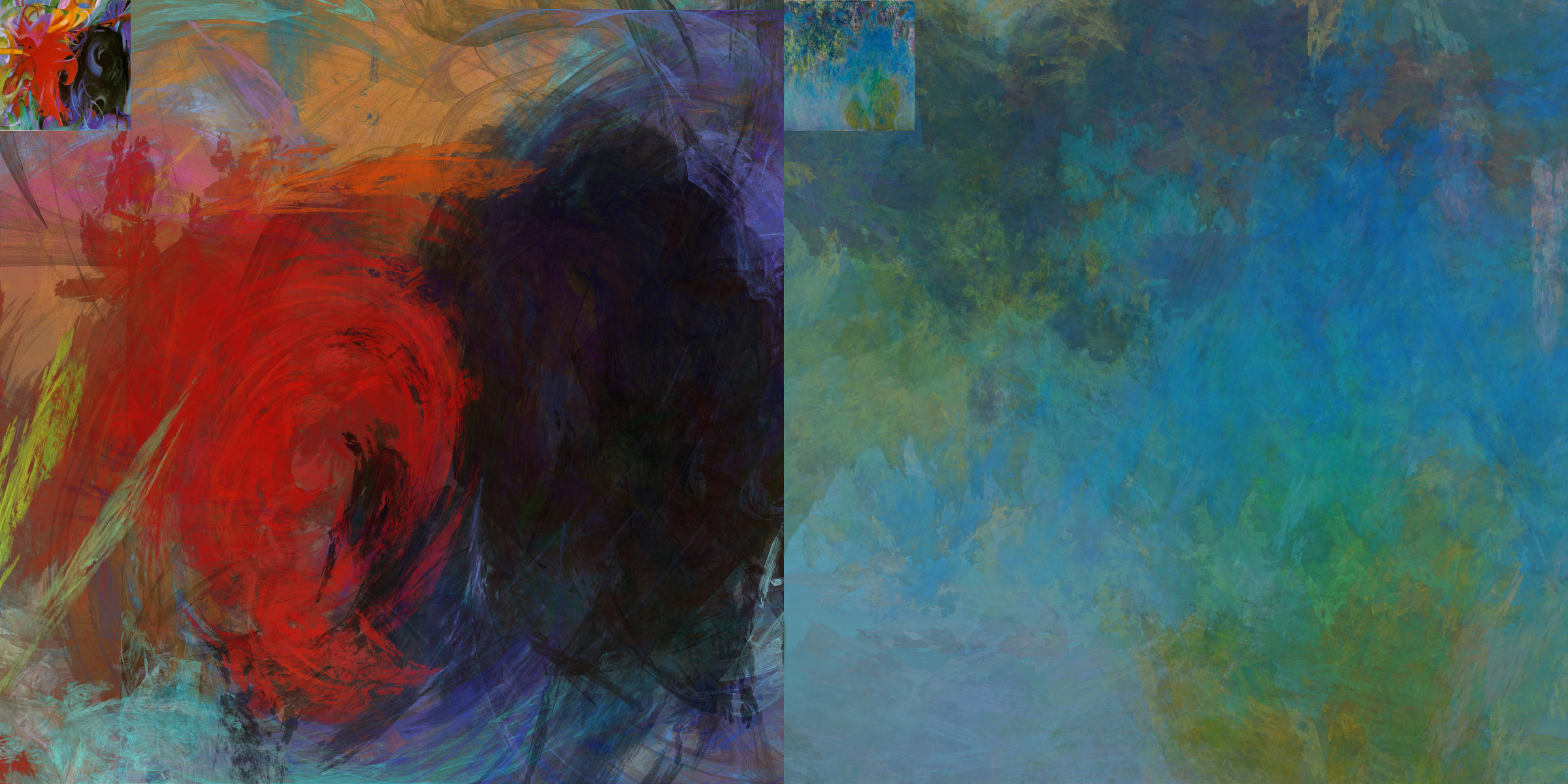}
    \caption{Two examples of fractal artwork created using the proposed approach. The reference images are displayed in the top left of each frame. \textbf{Left}: a composite fractal based on the painting \textit{Fighting Forms} by Franz Marc. \textbf{Right}: a composite fractal based on the painting \textit{Wisteria} by Claude Monet. }
    \label{fig:teaser}
    \Description{Two fractal artworks learned using our differentiable fractal rendering approach. The reference images are displayed in the top left of each frame.}
\end{teaserfigure}


\maketitle

\section{Introduction}

The term \textit{fractal} was first introduced by Benoit Mandelbrot in 1975 to describe a type of geometric pattern that he observed in his research \cite{mandelbrot}. Fractals are generally identified as such by the presence of self-similarity, meaning that sub-sections of a fractal will resemble the fractal in it's entirety. Fractals are also characterized by the presence of intricate detail when observed at many different scales. Mandelbrot was the first to discover that these patterns could be created from deceptively simple equations. The complex aesthetic features of fractal patterns were so compelling that the development of computer programs capable of rendering fractals promptly gave rise to a new genre of mathematical and geometric art. In the words of Mandelbrot: \textit{The source of fractal art resides in the recognition that very simple mathematical formulas that seem completely
barren may in fact be pregnant, so to speak, with an enormous amount of graphic structure} \cite{mandelbrot_art}.

While the mathematical formulas may be simple, the aesthetic landscape that fractal artists explore is certainly not. Artists create fractal artwork by configuring mathematical formulas, their parameters, and the fractal rendering algorithm according to their tastes and aesthetic objectives. However, it is difficult to know how one should alter a fractal rendering algorithm to achieve a particular outcome. Currently, fractal art creation primarily proceeds via trial and error, which is both non-intuitive and time consuming. Therefore, in this work, we study the problem of inverse fractal rendering. In other words, given a target image, we devise an algorithm capable of finding the latent parameters of a fractal such that the generated fractal resembles the target image. We believe that this approach is valuable as an artistic tool that makes fractal art creation faster and more intuitive.

\subsection{Fractal Flames}

Iterated function systems (IFS) are one of the many methods for creating fractals. IFS fractals were first conceived in 1981 by John E. Hutchinson \cite{hutchinson} and were brought into greater prominence by Michael Barnsley \cite{barnsley}. As suggested by the name, an IFS fractal is associated with a set of $N_F$ functions $F=\{f_i: X \rightarrow X\ | i = [1..N_F]\}$ called the generator functions. In this work, we will restrict ourselves to 2-dimensional fractals where $X = \mathbb{R}^2$. An IFS fractal can be rendered by running an algorithm called the chaos game, which proceeds as follows. First, an initial sample $s_0 \in X$ is selected. Next, for as many iterations as desired, a function $f \in F$ is selected at random and used to generate the next sample $s_{j} = f(s_{j-1})$. Finally, the sample positions $\{s_{0}, s_{1}, s_{2}, ...\}$ are plotted to create an image.

\begin{figure}
    \centering
    \includegraphics[width=\linewidth]{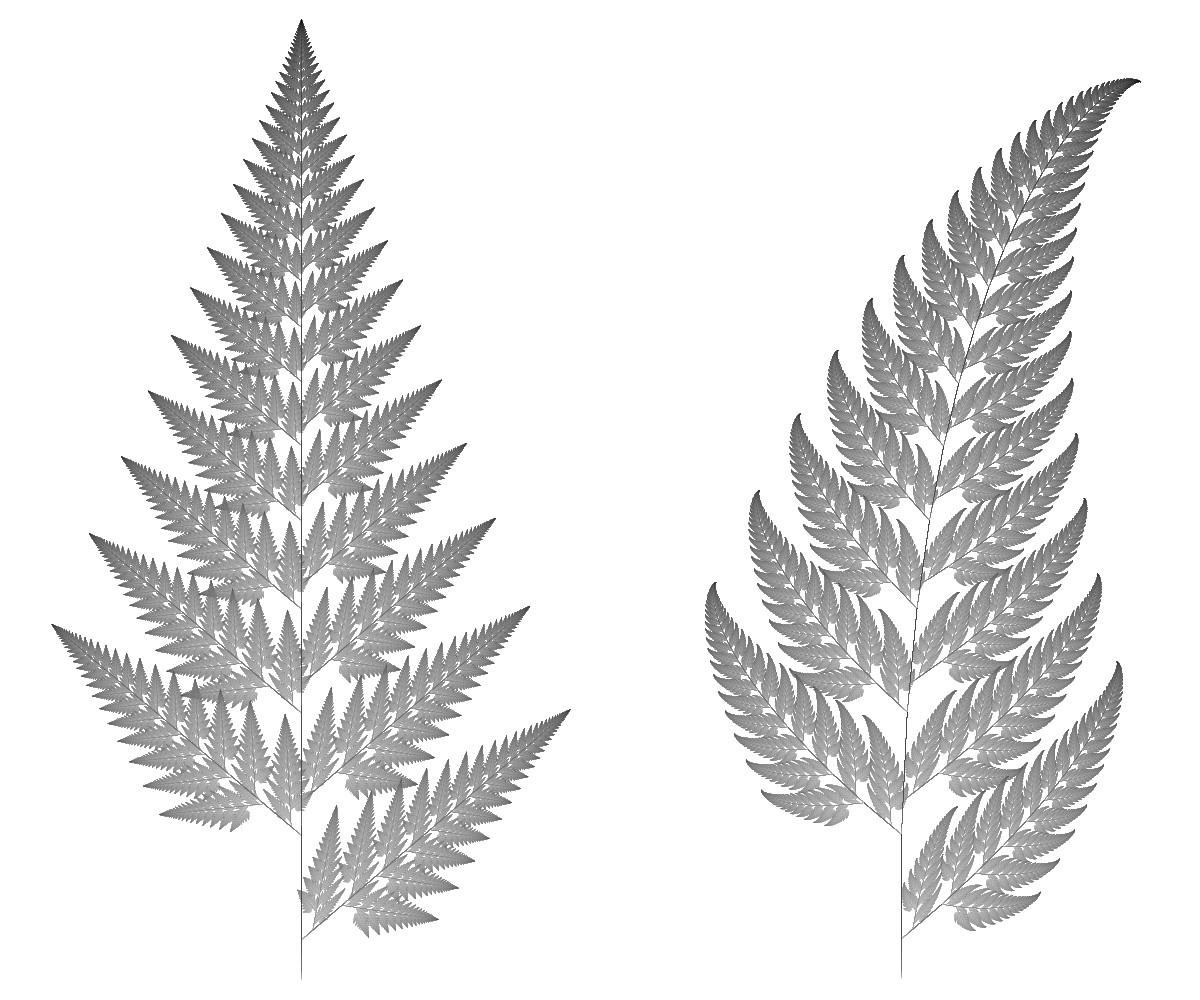}
    \caption{The Barnsley fern is an example of a linear IFS fractal. When the parameters of the generator functions are perturbed (left vs. right) the fractal structure is altered. In general, it is difficult to know which fractal parameters to modify, and how, in order to achieve a particular effect.}
    \label{fig:fern}
    \Description{The Barnsley fern IFS fractal rendered with two different sets of parameters.}
\end{figure}

The first IFS fractals used linear generator functions and produced binary or grayscale images (see Figure \ref{fig:fern}). In 1992, Scott Draves introduced the fractal flame algorithm which greatly expanded the aesthetic range of IFS fractals \cite{draves_ff, draves_ff2}. The three core ideas introduced by the fractal flame algorithm are: 1) the use of non-linear generator functions, the different varieties of which produce distinctive visual effects (see Figure \ref{fig:variations}), 2) a log-density tone mapping approach, which reveals additional detail in fractal images, and 3) a method of coloring samples based on the most recent generator functions from which they were produced. 

Subsequently, several software programs were developed for creating and editing fractal flame art, including Electric Sheep \cite{electric_sheep}, JWildfire \cite{jwildfire}, Apophysis \cite{apophysis}, and Chaotica \cite{chaotica}. Fractal flames are now a widely used approach for creating fractal art.

\begin{figure*}
    \centering
    \includegraphics[width=\linewidth]{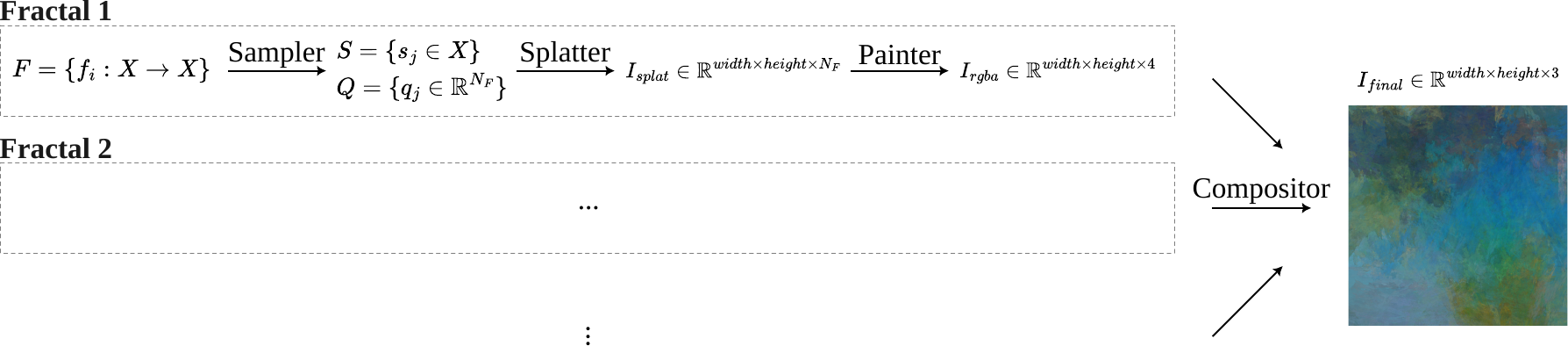}
    \caption{A flow diagram of our differentiable fractal rendering pipeline. First, a sampler uses a set of generator functions to produce arrays of sample positions and quality vectors. Second, a splatter splats each sample quality vector onto an image buffer at the corresponding sample position. Third, a painter maps the vector at each splat buffer pixel to an RGBA value. Finally, the compositor composites RGBA buffers from one or more fractals over a background color to produce the final image. The pipeline is end-to-end differentiable.}
    \label{fig:diagram}
    \Description{A flow diagram of our differentiable fractal rendering pipeline.}
\end{figure*}

\subsection{Inverse Fractal Rendering}

Several different methods have been proposed to configure and control fractal generation. Recently, \cite{portal} introduced a method of creating self-similarity in arbitrary 2D and 3D shapes. This follows a line of work that seeks to enable artistic control over fractal geometry
\cite{kim1, kim2, kim3}. The class of fractals investigated in these works, sometimes called escape time fractals, are distinct from IFS fractals in that they categorize points based on whether or not their magnitude diverges or "escapes" to infinity upon repeated application of a function. Furthermore, these approaches seek only to optimize fractal geometry, whereas our approach optimizes both geometry and color. 

Several approaches have been specifically developed to infer the parameters of IFS fractals \cite{lankhorst1995iterated, guiterrez, vrscay1989iterated, vrscay1991iterated, kya2001optimization}. These approaches primarily make use of genetic algorithms or moment matching, and were not applied for the purpose of fractal art creation. In fact, much of the initial research into inverse IFS fractal rendering was directed towards the application of image compression \cite{fractal_comp}, where the goal is to recreate an image as accurately as possible while minimizing the size of the latent or compressed representation of the image. This objective is distinct from that of the fractal flame approach, which seeks to create aesthetically interesting fractal artwork. The approach proposed in this work combines elements of both image compression and artistic creation. The algorithm seeks to recreate a provided image, however the objective is not compression, but rather the creation of aesthetically interesting fractal compression artifacts. 

Recently, \cite{tu} presented the first approach for learning IFS fractal parameters by gradient descent optimization. Their core contribution was to propose a differentiable method for rendering point samples produced by the chaos game. The classic approach to rendering an IFS sample is to increment the pixel within which the sample falls by some fixed amount. Instead, they increment pixels in a neighborhood of the sample position according to the distance from the sample to each pixel center, weighted by a radial basis function. In this approach, the fractal image pixel values are continuously related to the sample positions, thereby enabling gradient backpropagation. They liken their approach to the differentiable soft quantization module of \cite{qian} and we note that it also bears similarity to splatting operations found in some work on differentiable triangle rasterization \cite{cole}. 

The approach demonstrated by \cite{tu} has several important limitations as a tool for creating fractal art. First, the approach uses only linear generator functions. As demonstrated by the many pieces of artwork created with the fractal flame algorithm \cite{draves_ff}, and as can be seen directly in Figure \ref{fig:variations}, the use of non-linear generator functions greatly expands the aesthetic range of IFS fractals. Second, color images are not supported in their approach. Finally, their approach struggles to represent complex images.

\subsection{Contributions}

In this work, we extend the differentiable rendering approach of \cite{tu} and address the limitations described above. Our approach can learn colorful IFS fractals with non-linear generator functions. Furthermore, we demonstrate the use of a differentiable compositor to combine multiple fractal flames for more representational flexibility. 

\section{Overview and Implementation}

A high level flow diagram of our differentiable rendering pipeline is shown in Figure \ref{fig:diagram}. There are four main components. In this section, each component will be described, including inputs, outputs, parameters and other important considerations. Our implementation relies on the Taichi language \cite{taichi}, which supports both automatic differentiation and GPU acceleration. The code is publicly available\footnote{
\href{https://github.com/JJBannister/LearnableFractalFlames}{\textcolor{blue}{github.com/JJBannister/LearnableFractalFlames}}}.

\subsection{Sampler}
The role of a sampler is to run the chaos game algorithm. Therefore, each sampler is associated with a set of $N_F$ generator functions $F = \{f_i: X \rightarrow X | i = [1.. N_F]\}$ with trainable parameters $\theta_i$. The number of generators and the structure of each generator function is defined when constructing the sampler. The functions in $F$ may be non-linear, but must be differentiable.

In the original chaos game algorithm, the generator functions are randomly sampled in each iteration. In a differentiable rendering approach, we must slightly re-organize this process to ensure differentiability. Following \cite{tu}, we sample the generator order separately from the generation of samples, similar to the re-parameterization trick used in variational auto-encoders \cite{kingma}. Thus, the random sampling is treated as constant during back-propagation and the gradient flow circumvents the non-differentiable stochastic process. The generator order is, therefore, an array of indices $G = \{g_j \in [1..N_F] | j = [1..N_S]\}$ where $N_S$ is the number of samples generated. 

One output of the sampler is a set of $N_S$ sample positions $S=\{s_j \in X | j = [1..N_S]\}$. The sample positions are computed by serially applying the generator functions in the order defined by the array $G$ (Equation \ref{eq:positions}). 

\begin{equation}\label{eq:positions}
    s_{j} = f_{g_j}(s_{j-1})
\end{equation}

A second output of the sampler is a set of sample quality vectors $Q = \{q_j \in \mathbb{R}^{N_F}| j = [1..N_S]\}$ in correspondence with the sample positions. The sample quality vectors (calculated using Equation \ref{eq:qualities}) reflect which functions have most recently acted upon the sample. 

\begin{equation}\label{eq:qualities}
    q_j = \sum_{k=0}^{j} \beta^{k-j}  e_{g_k}  
\end{equation}

Here, $e_{g_k}$ represents a one-hot vector $\in \mathbb{R}^{N_F} $ with a one at the index corresponding to the generator index $g_k$ of sample $k$. The learnable parameter $\beta$ represents the quality decay parameter, which controls how quickly the influence of a generator function disappears from the sample quality vector as more generator functions are applied to the sample. Parameter $\beta$ is learnable and is constrained to be greater than 1.

\begin{figure*}
    \centering
    \includegraphics[width=\linewidth]{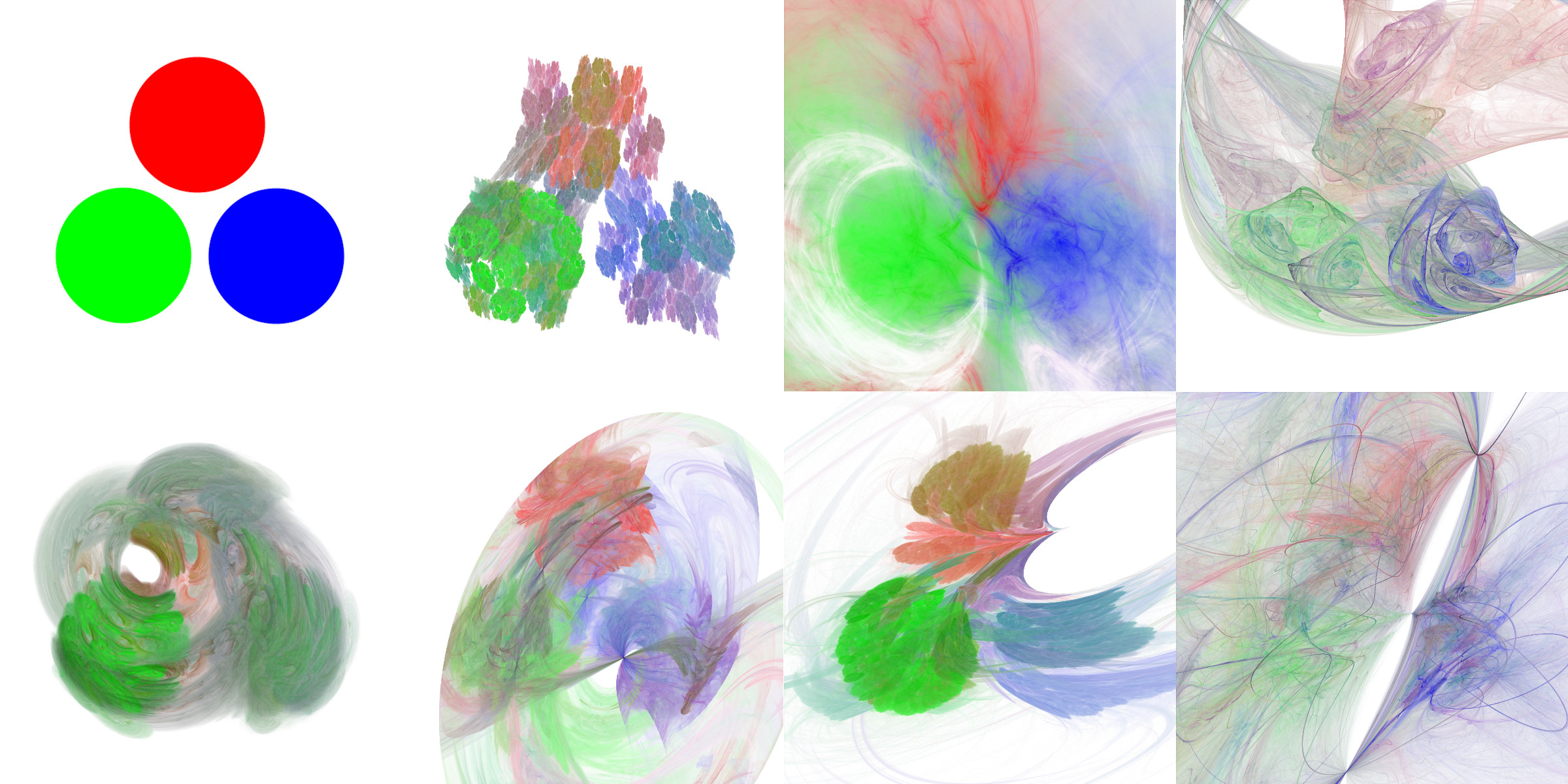}
    \caption{Fractal flames learned from a simple reference image. \textbf{Top row} (left to right): the reference image, a linear variation, a spherical variation, a hankerchief variation. \textbf{Bottom row} (left to right): an exponential variation, a disk variation, a heart variation, a power variation.}
    \Description{Seven fractal flames learned from a simple reference image consisting of three colorful circles.}
    \label{fig:variations}
\end{figure*}

\subsection{Splatter}
The role of a splatter is to take the outputs of a sampler, and to splat the sample quality vectors onto an image buffer at positions defined by the sample positions. The output of the splatter is, therefore, a 2D image buffer where each pixel has $N_F$ channels. A splatter has learnable parameters that correspond the the final transform described in \cite{draves_ff}. This linear transform is applied to all sample positions before they are mapped to pixel space coordinates. After a sample position has been mapped to pixel space coordinates, a center pixel (within whose bounds the sample position falls) is identified, and the sample is splatted in the 3x3 pixel neighborhood of the center pixel. Each pixel in the 3x3 neighborhood is incremented by the quality vector of the sample multiplied by a weight. The weight of a sample at a pixel whose center is $d$ pixels distant from the sample is $e^{-2d^2}$. This corresponds to an isotropic Gaussian kernel with standard deviation $\sigma=0.5$ pixels. This splatting approach follows the same general idea as \cite{tu} and 
\cite{cole}. The splatted pixel values are continuously related to the sample positions, thereby allowing gradient back-propagation.

\subsection{Painter}
The role of a painter is to transform the vector $q_{xy} \in \mathbb{R}^{N_F}$ at each pixel of the splat buffer into an RGBA pixel $p_{xy} \in \mathbb{R}^{4}$. A painter is associated with one RGBA parameter for each generator function $C = \{c_i \in \mathbb{R}^4 | i = [0..N_F]\} $. First, for each pixel in the splat buffer, the splat weight $w_{xy} = sum(q_{xy})$ is computed, and the maximum value $w_{\text{max}}$ across the splat buffer is stored. Next, for each pixel, a weight alpha $\alpha_{xy} = \log(w_{xy}) \div \log(w_{\text{max}})$ is computed. Finally, the RGBA value $p_{xy}$ of each pixel is computed using Equation \ref{eq:rgba}.

\begin{equation}\label{eq:rgba}
    p_{xy} = \frac{\alpha_{xy}}{w_{xy}}   \sum_{i=0}^{N_F} c_i \cdot q_{{xy}_i}  
\end{equation}

There are two sources of transparency in this approach. The first is the learnable alpha values of parameters in $C$. The second is the splat weight alpha $\alpha_{xy}$. We also note that this approach, based on the log-density approach described in \cite{draves_ff}, is simply an ad-hoc tone mapping method with learnable parameters. Any differentiable function of the form $\mathbb{R}^{N_F} \rightarrow \mathbb{R}^{4}$, which outputs an RGBA value could be applied here. 

\subsection{Compositor}
The final module in our differentiable fractal rendering pipeline is a differentiable compositor. A compositor is associated with a learnable RGB background color parameter. The role of the compositor is to composite one or more RGBA buffers over the background color. There is only one compositor in any pipeline, while there may be one or more of the other components (see Figure \ref{fig:diagram}).

\subsection{Training and Evaluation}
Different considerations apply when optimizing fractal parameters compared to generating a final image. When training, speed of execution is critical. When generating a final image, the quality of the output is most important, and it is acceptable to increase runtime. Therefore, we set hyper-parameters to control the quality-speed trade-off separately for each scenario. When training, we use fewer samples and lower image resolution. After learning the fractal parameters, we generate a full resolution image using many more samples to ensure a high quality result.

\section{Experiments and Results}
In our experiments, we aim to demonstrate that our differentiable rendering approach allows for controllable fractal art creation. Furthermore, we investigate several aspects of the optimization process, including parameter initialization and hyper-parameter selection, in order to understand the aesthetic effects that they have on the generated fractal art.  

All experiments were performed using an Nvidia 2070 Super GPU. The most computationally expensive experiments, in which multi-flame fractals are trained to emulate paintings, complete in roughly 20 minutes. The simpler experiments complete in one or two minutes. For all experiments, the number of samples per flame used during training was $10^6$, and during evaluation $10^{10}$. A mean squared error loss function and a simple gradient descent optimizer were used in all experiments.

\subsection{Learning Non-Linear Fractal Flames}

In Scott Draves' description of the fractal flame algorithm \cite{draves_ff}, 48 different non-linear generator function variations were introduced. Each of these variations can be combined with others in an infinite number of different ways to create different aesthetic effects. In our formulation, each generator need only be a differentiable function $f: X \rightarrow X$, which leaves considerable room for experimentation and exploration. Accepting that a comprehensive study of this space is impossible, we first demonstrate our approach using a small selection of non-linear variations.

Figure \ref{fig:variations} shows a simple reference image alongside optimized fractals representing a linear variation and 6 different non-linear variations. Many of these fractals are not capable of capturing the structure of the reference image closely. This is especially true when the intrinsic symmetry of the fractal variation does not match the reference image. Nevertheless, they all show a tendency to assign color and density to appropriate areas of the image while exhibiting distinct aesthetic qualities associated with the different variations. In these experiments, the background color of the image was set to white and was not learned. Each fractal was trained at a resolution of $200\times200$ pixels and has 8 generator functions.

\subsection{Exploring the Learning Process}

\begin{figure}
    \centering
    \includegraphics[width=\linewidth]{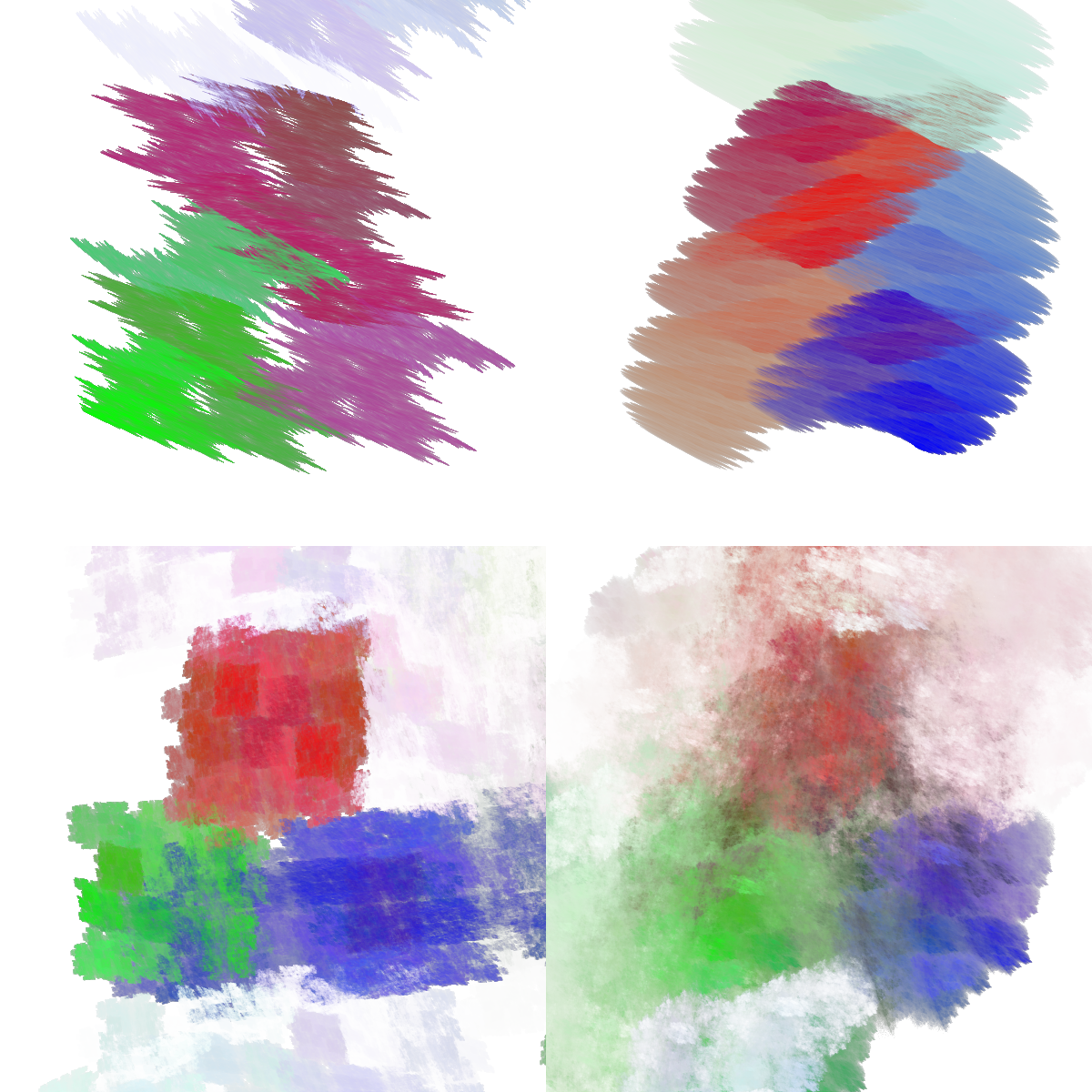}
    \caption{Top row: two fractal flames with 4 generator functions. Bottom row: two fractal flames with 16 generator functions. Fractals with fewer generator functions generally have simpler geometry and color palettes. All fractals are linear and were trained on the reference image shown in Figure \ref{fig:variations}.}
    \Description{}
    \label{fig:n_generators}
\end{figure}

During our exploration and experimentation with fractal flame optimization pipelines, we observed several phenomena that had important aesthetic impacts on the fractal generation process. We experimented with the number of generator functions used for each fractal. The results in Figure \ref{fig:n_generators} show that increasing the number of generator functions in a fractal, increases the ability of the fractal to represent more complex patterns. This result is fairly predictable, yet important. Increasing the number of generator functions means that there are more free parameters in the model, with respect to both geometric structure and color.

\begin{figure*}
    \centering
    \includegraphics[width=\linewidth]{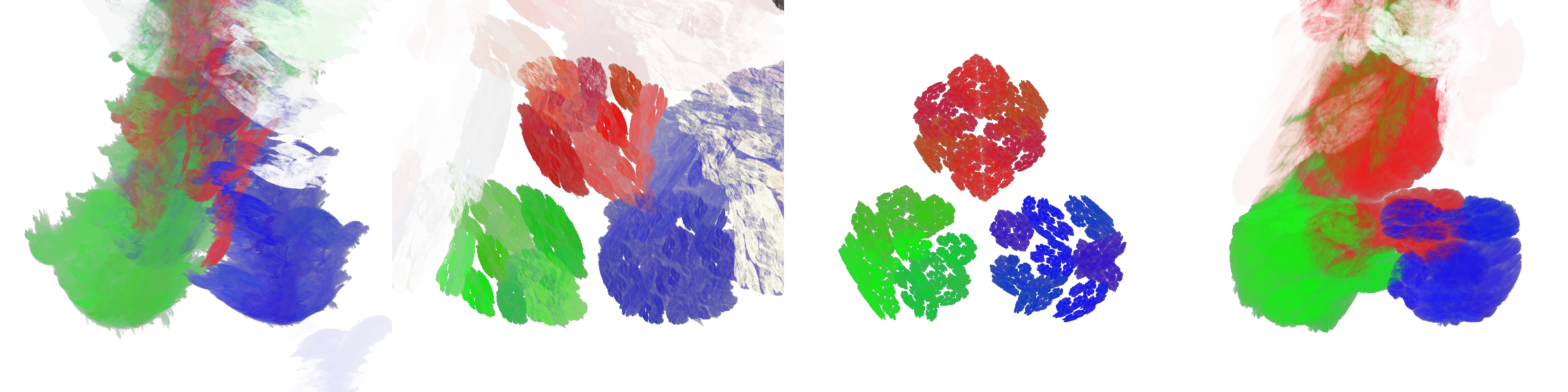}
    \caption{Four fractal flames with 8 generator functions that were trained using different initial parameter values. When the initial parameters are different, the optimization process will often find a different local minimum, rather that a common global minimum. All fractals are linear and were trained on the reference image shown in Figure \ref{fig:variations}.}
    \Description{}
    \label{fig:random_seed}
\end{figure*}

Another important aspect of the learning algorithm is a high degree of sensitivity to initial conditions. This means that choosing different initial values for the fractal parameters (prior to running the optimization process) results in different final fractal parameters and visually different results. Figure \ref{fig:random_seed} shows several outputs from fractal optimization processes that differed only in that they used different random seeds to initialize the fractal parameters. These results indicate that the loss landscape associated with the optimization process is non-convex and, therefore, that the optimization processes are finding different local minima rather that a common global minimum. This fact is highly relevant to the artistic application of this optimization process, as it means that artists may achieve different outcomes by manually configuring the initial parameters or by initializing with different random seeds. 

\begin{figure}
    \centering
    \includegraphics[width=\linewidth]{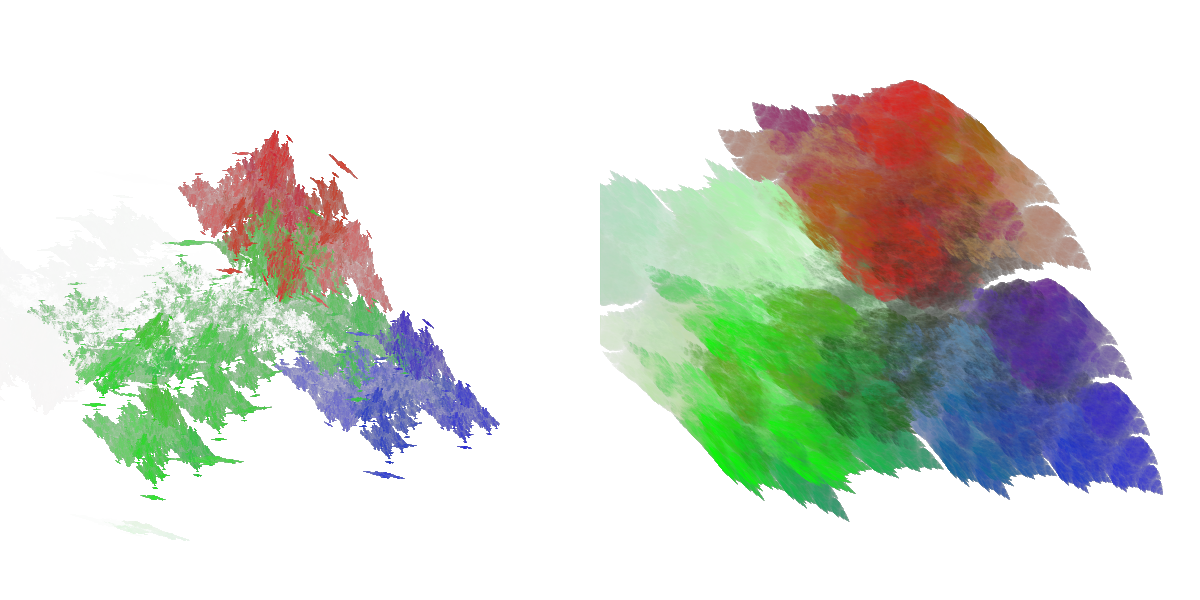}
    \caption{Top: a linear fractal (left) and non-linear spherical fractal (right) trained at a resolution of $20\times20$. Bottom: a linear fractal (left) and non-linear spherical fractal (right) trained at a resolution of $400\times400$. Fractals trained at lower resolutions tend to have more thin and sharp structures. All fractals used 8 generator functions and were trained on the reference image shown in Figure \ref{fig:variations}.}
    \Description{Fractal flames that were trained on the same reference images as Figure \ref{fig:teaser} but with different generator function variations.}
    \label{fig:resolution}
\end{figure}

Another interesting aspect of fractal optimization that we observed was that the resolution of the image used during training had a strong effect on the outcome of the optimization process. The splatting operation that enables the fractal rendering process to become differentiable, applies a Gaussian blur kernel with a standard deviation of 0.5 pixels to the resulting image. Because the magnitude of the blur is a function of the pixel size, reducing image resolution results in a larger blurring kernel. A larger blurring kernel affects the gradients that are used in the optimization process, which results in different outcomes. Figure \ref{fig:resolution} shows several outputs from fractal optimization processes that differed only in that they used different image resolutions when optimizing the fractals. The effect of using smaller image resolutions was generally to produce fractals with more thin and sharp structures.

\subsection{Multi-Fractal Compositions}

Finally, we trained fractals with the objective of creating complex visual art. We selected two famous paintings in the public domain to use as reference images for these experiments (\textit{Fighting Forms} by Franz Marc \cite{fighting_forms}, and \textit{Wisteria} by Claude Monet \cite{wisteria}). All experiments discussed in this section used a training image resolution of $200\times200$ pixels. 

\begin{figure*}
    \centering
    \includegraphics[width=\linewidth]{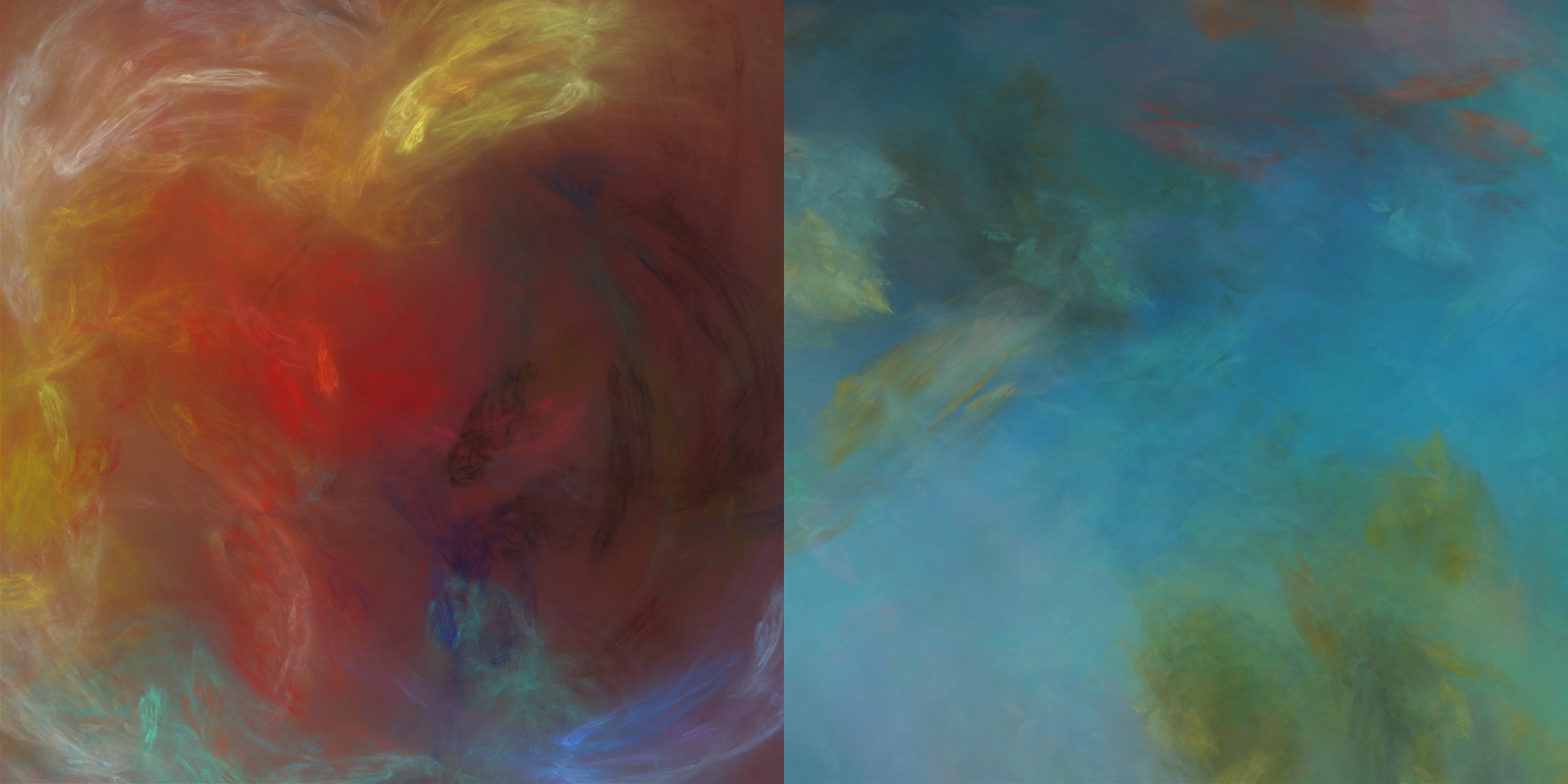}
    \caption{A non-linear spherical (left) and a linear (right) fractal flame trained on the same reference images as Figure \ref{fig:teaser}. These experiments used only a single fractal, unlike those in Figure \ref{fig:teaser} which composited multiple fractals.}
    \Description{Individual fractal flames that were trained on the same reference images as Figure \ref{fig:teaser}.}
    \label{fig:single_flame}
\end{figure*}

Our initial experiments used a single fractal flame with 24 generator functions for each image. Some of these results are shown in Figure \ref{fig:single_flame}. With the goal of increasing the complexity of the generated fractal artwork, we began experimenting with multi-fractal compositions. Figure \ref{fig:teaser} shows two results from these experiments. For \textit{Wisteria}, we composed 4 linear fractal flames with 24 generator functions each. For \textit{Fighting Forms}, we composed 3 non-linear fractal flames, which included spherical, handkerchief, and exponential variations. In this experiment, we found that our pipeline was capable of capturing the gross structure of the reference image quite closely, while still exhibiting the desirable fractal aesthetic qualities.

\section{Limitations}
The differentiable fractal rendering approach described here has several limitations. One limitation of our approach is the potential for training instability when using non-linear variations. In practice, we found that this was often not an issue. However, we did not explore all possible variations and stability issues may still be encountered in some scenarios. 

A limitation with our current fractal flame implementation is a lack of support for stochastic generator functions. In addition to using random number generation when determining the generator function order, some of the fractal flame variations also include random number generation in the generator functions themselves \cite{draves_ff}. Because differentiating through stochastic sampling is not possible, supporting such flame variations would require another application of the re-parameterization trick. 

\section{Conclusion and Future Work}
In summary, we developed a state-of-the-art differentiable IFS fractal rendering approach that supports non-linear generator functions, color images, and multi-fractal compositions. Differentiable fractal rendering in combination with gradient descent optimization allow artists to use reference images in order to control the generation of colorful and complex fractal flame artwork. An exciting avenue for future work would be to integrate this approach into interactive graphical applications similar to existing fractal flame creation software \cite{chaotica, apophysis, electric_sheep, jwildfire}. This would allow fractal artists without a background in programming to access and explore this technology.


\bibliographystyle{ACM-Reference-Format}
\bibliography{bibliography}

\end{document}